\documentclass[12pt]{article}
\usepackage{graphicx}  
\usepackage{latexsym}
\usepackage{amsbsy}
\usepackage{sectsty}
\usepackage{cite}

\sectionfont{\large}

\newcommand{\kB}{k_{\mathrm{B}}}

\begin{document}

\begin{center}
  {\Large \bf Simulation of Magnetization Switching in Nanoparticle Systems}

\vspace{2cm}
  
{\large
  D.\ Hinzke and U.\ Nowak\\
  Theoretische Physik, Gerhard-Mercator-Universit\"{a}t\\
  47048 Duisburg, Germany}
\end{center}

\vspace{2cm}

\noindent
Pacs-numbers: 75.10.Hk; 75.40.Mg; 75.40.Gb

\section*{Abstract}
Magnetization reversal in magnetic nanostructures is investigated
numerically over time-scales ranging from fast switching processes on
a picosecond scale to thermally activated reversal on a microsecond
time-scale. A simulation of the stochastic Landau-Lifshitz equation of
motion is used as well as a time quantified Monte Carlo method for the
simulation of classical spin systems modeling magnetic Co
nanoparticles. For field pulses larger than the Stoner-Wohlfarth limit
spin precession effects govern the reversal behavior of the particle
while for lower fields a magnetization reversal is only possible when
it is assisted by thermal fluctuations.

\newpage

\section{Introduction}
The miniaturization of magnetic structures plays an important role for
fundamental research as well as for technical applications.  This
leads to an incremental interest in the understanding of the behavior
of small magnetic particles and structures down to the nanometer
scale.  But with decreasing size of the magnetic system thermal
activation becomes relevant.  Hence, much effort is focused now on the
understanding of magnetization dynamics at finite temperatures.
\cite{nowakARCP00}.

In the following, we briefly describe numerical techniques for the
study of magnetization dynamics in nanostructures, modeled as
classical spin systems, where a finite temperature is taken into
account. First, we will focus on the underlying model and the two
basic methods, Langevin dynamics \cite{lyberatosJAP93} and Monte Carlo
simulation \cite{nowakPRL00}. Then we study the magnetization reversal
in Co nanoparticles.  Starting with the deterministic spin dynamics on
short time scales which plays a crucial role in high-speed data
storage \cite{backPRL98,bauerJAP00,leineweberPB00}, we go on to the
probabilistic long-time behavior where a thermally assisted reversal
can occur even for magnetic fields below the coercive field
\cite{brownPR63,braunPRL93,wernsdorferPRL97,hinzkePRB00,hinzkeJMMM00}.


\section{Classical Spin Model}
\label{s:model}

The micromagnetic properties of a nanoparticle can be described using
a model of classical magnetic moments which are localized on a given
lattice.  Such a spin model can be motivated following different
lines: on the one hand it is the classical limit of a quantum
mechanical, localized spin model --- the Heisenberg model
\cite{aharoniBOOK96}. On the other hand, a classical spin model can
also be interpreted as the discretized version of a micromagnetic
continuum model \cite{schreflEM01}, where the charge distribution for
a single cell of the discretized lattice is approximated by a point
dipole \cite{aharoniBOOK96,hubertBOOK98}.

The interpretation as an atomic model restricts the use of computer
simulations to the investigation of rather small systems of only a few
million atoms - corresponding to particle sizes of only a few
nanometers. On the other hand, within a continuum model, the space
might be discretized on a much larger length scale, as compared to an
atomic distance.  However, in continuum theory usually a constant
absolute value of the magnetization vector is assumed, an assumption
which fails for higher temperatures since the space averaged
magnetization breaks down when approaching the critical temperature.
Hence, one expects correct thermal properties only in the limit of
small cell sizes of the order of atomic distances.

In the following, let us consider a classical three dimensional
Heisenberg Hamiltonian for localized spins,
\begin{eqnarray}
  \label{e:ham}
  \textstyle
 {\cal H} &=& - J \sum_{\langle ij \rangle} {\mathbf S}_i \cdot {\mathbf
    S}_j - \mu_s{\mathbf B}\cdot \sum_i {\mathbf S}_i -
    d_{z} \sum_i  (S_{i}^{z})^2 \nonumber \\ &&
    - w \sum_{i<j} \frac{3({\mathbf S}_i \cdot {\mathbf e}_{ij})({\mathbf
      e}_{ij} \cdot {\mathbf S}_j) - {\mathbf S}_i \cdot {\mathbf
      S}_j}{r^3_{ij}},
\end{eqnarray}
where the ${\mathbf S}_i = {\boldsymbol \mu}_i/\mu_s$ are three
dimensional magnetic moments of unit length representing atomic
magnetic moments. The first sum is the ferromagnetic exchange of the
moments with the coupling constant $J$.  The second sum is the
coupling of the magnetic moments to an external magnetic field $B$,
the third sum represents a uniaxial anisotropy, here, for $d_z > 0$
favoring the $z$ axis as easy axis of the system, and the last sum is
the dipolar interaction where $w = \mu_0 \mu_s^2 /(4 \pi a^3)$
describes the strength of the dipole-dipole interaction.  The
${\mathbf e}_{ij}$ are unit vectors pointing from lattice site $i$ to
$j$ and $r_{ij}$ is the distance between these lattice sites in units
of $a$.  The dipole-dipole interaction can be computed efficiently
with the help of fast Fourier transformation methods
\cite{yuanIEEE92,berkovPSS93}.  One should however note that in a
Monte Carlo simulation with a single-spin flip algorithm the FFT
method is an approximation the implementation of which was described
in details before \cite{hinzkeJMMM00}.

The transformation of the above introduced atomic parameters to the
material parameters which are usually used in a continuum model is
given by $J = 2aA_{x}$ where $A_{x}$ is the exchange energy, $d_{z} =
Ka^{3}$ where $K$ is the anisotropy energy density, and $\mu_{s} =
M_{s}a^{3}$ where $M_{s}$ is the spontaneous magnetization.

\section{Landau-Lifshitz Equation and Langevin Dynamics}

In the short time limit spin precession is important which can be
taken care of by studying the corresponding equation of motion. The
basic numerical approach which includes thermal activation, is the
direct numerical integration of the Langevin equation of the problem.
In order to obtain thermal averages one has to calculate many of these
trajectories starting with the same initial conditions taking an
average over these trajectories for the quantities of interest.  This
method is referred to as the Langevin dynamics formalism
\cite{lyberatosJAP93}.

The underlying equation of motion for a magnetic system is the
Landau-Lifshitz-Gilbert (LLG) equation,
\begin{equation}
\frac{\partial {\mathbf S}_i}{\partial t} =
  -   \frac{\gamma}{(1+\alpha^2)\mu_s}
  {\mathbf S}_i \times \Big[{\mathbf H}_i(t) +
  \alpha \big ({\mathbf S}_i \times {\mathbf H}_i(t) \big) \Big],
\label{e:llg}
\end{equation}
with the gyromagnetic ratio $\gamma = 1.76 \times 10^{11} ({\mathrm
  Ts})^{-1}$, the dimensionless damping constant $\alpha$, and the
internal field $ {\mathbf H}_i(t) = {\boldsymbol \zeta}_i(t) -
\partial {\cal H} /\partial {\mathbf S}_i$. Langevin
dynamics is introduced here in form of the noise ${\boldsymbol \zeta}_i(t)$
which represents thermal fluctuations, with $\langle {\boldsymbol
  \zeta}_i(t) \rangle = 0$ and $ \langle \zeta_i^{\eta}(t)
\zeta_j^{\theta}(t') \rangle = 2 \delta_{ij} \delta_{\eta \theta}
\delta(t-t') \alpha \kB T \mu_{s}/ \gamma$ where $i, j$ denote once
again lattice sites and $\eta, \theta$ Cartesian components.

The LLG equation with Langevin dynamics is a stochastic differential
equation with multiplicative noise. For this kind of differential
equation a problem arises which is called the It\^{o}-Stratonovich
dilemma \cite{GreinerJSP88}. As a consequence, different time
discretization schemes may converge to different results with
decreasing time step. As was pointed out in \cite{garciaPRB98} the
multiplicative noise in the Langevin equation above has to be treated
by means of the Stratonovich interpretation.  An appropriate
discretization scheme leading to a Stratonovich interpretation is the
Heun method \cite{GreinerJSP88,garciaPRB98,nowakARCP00} which is used
in the following.

\section{Monte Carlo Methods}

In the long time limit only spin relaxation and thermal fluctuations
are relevant which can be studied very conveniently using Monte Carlo
methods with quantified time step \cite{nowakPRL00}.  Within a Monte
Carlo approach \cite{binderBOOK97} trajectories in phase space are
calculated following a master equation \cite{reifBOOK65} for the time
development of the probability distribution $P_s(t)$ in phase space,
\begin{equation}
  \label{e:master}
  \frac{\mbox{d} P_s}{\mbox{d} t} = \sum_{s'} (P_{s'} w_{s' \to s} -
    P_s w_{s \to s'}).
\end{equation}
Here, $s$ and $s'$ denote different states of the system and 
$w_{s' \to s}$ is the transition rate for a change from a state $s'$
to a state $s$. These rates have to fulfill the condition
\cite{reifBOOK65}
\begin{equation}
  \label{e:det-bal}
  \frac{w_{s \to s'}}{w_{s' \to s}} = \exp\Big[\frac{E(S) - E(S')}{\kB
    T}\Big].
\end{equation}
The master equation describes exclusively the coupling of the system
to the heat bath \cite{reifBOOK65}. Hence, only the irreversible part
of the dynamics of the system is considered including only the
relaxation and the fluctuations. A Monte Carlo simulation does not
include the energy conserving part of the equation of motion. Hence, no
precession of magnetic moments will be found.

Monte Carlo approaches in general have no physical time associated
with each step of the algorithm. However, recently a time quantified
Monte Carlo method was proposed in \cite{nowakPRL00} and later
successfully applied to different model systems
\cite{hinzkePRB00,hinzkeJMMM00,smirnovJAP00}.  Here, the
interpretation of a Monte Carlo step as a realistic time interval
$\Delta t$ was achieved by a comparison of one step of the Monte Carlo
process with a time interval of the LLG equation in the high damping
limit. We will use this algorithm in the following.  The trial step of
this algorithm is a random movement of the magnetic moment within a
cone with a given size $r$ with
\begin{equation}
  r^2 = \frac{20 k_B T \alpha \gamma}{(1+\alpha^2) \mu_s} \Delta t.
  \label{e:trial}
\end{equation}
In order to achieve this efficiently one constructs a random vector with
constant probability distribution within a sphere of radius $r$. This
random vector is added to the initial moment and subsequently the
resulting vector is normalized \cite{nowakPRL00}.

Using this algorithm one Monte Carlo step represents a given time
interval $\Delta t$ of the LLG equation in the high damping limit as
long as $\Delta t$ is chosen appropriately (for details see
\cite{nowakARCP00}).

\section{Precessional Reversal in Co Nanoparticles}
\label{s:results}

In the following we consider Co nanoparticles, where the material
parameters are $A_x = 1.3 \cdot 10^{-11}$ J/m, $K = 6.8 \cdot 10^{5} $
J/m$^2$ and $M_s = 1.43 \cdot 10^6$ A/m. For simplicity we simulate a
simple cubic lattice with atomic distance $a = 0.25$ nm. Our
simulation starts with a spin configuration where all magnetic moments
point into the $z$ direction, aligned with the easy axis, and with the
$z$ component of the external magnetic field antiparallel to the
magnetization so that the system is in an unstable, or at least
metastable state.

In sufficiently small particles the magnetic moments rotate coherently
during the magnetization reversal. A quantitative description of
coherent rotation in ellipsoidal single domain particles was developed
by Stoner and Wohlfarth \cite{stonerPTRS49}. Depending on the angle
between the applied field $\mathbf{B}$ and the $z$ (easy) axis of the
system, the coercive field $B_c$ varies following the so-called
Stoner-Wohlfarth asteroid \cite{stonerPTRS49}. Under an angle of
$45^{\circ}$ it is $B_c = d_{\mathrm{eff}}V /\mu_{\mathrm s}$ where
$V$ is the volume of the particle and $d_{\mathrm{eff}}$ is an
effective anisotropy constant.  In our case it is $B_c \approx 0.7$ T.

\begin{figure}
  \includegraphics[width=.85\textwidth, bb = 100 260 450
  470]{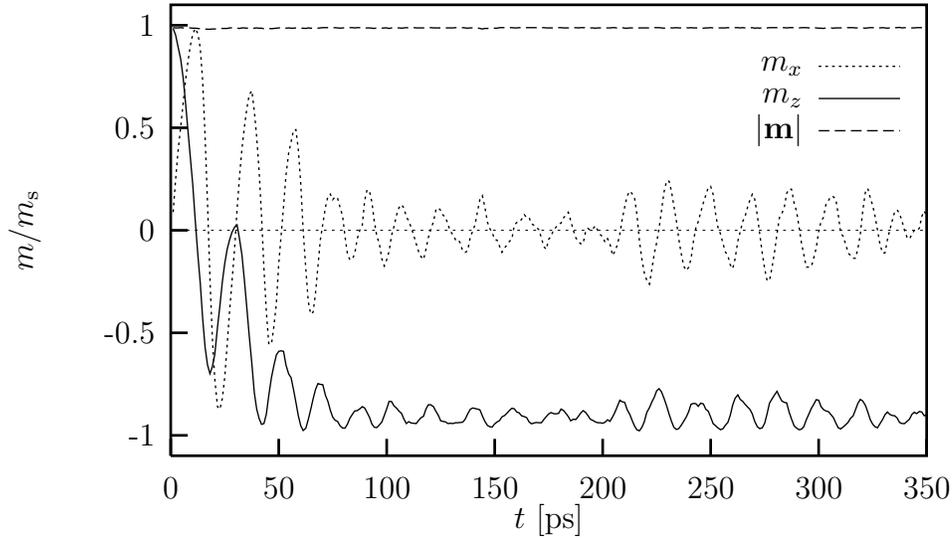}
  \caption{Reduced magnetization vs. time for a Co nanoparticle. The
    data are from Langevin dynamics simulations. $B = 1.13$ T and $T =
    16$ K.}
  \label{f:mag_LD}
\end{figure}

First, we are interested in fast switching processes where the applied
field is higher than the coercive field and the reversal is dominated
by spin precession. We simulate the system with Langevin dynamics as
described before. Figure \ref{f:mag_LD} shows the time dependence of
the magnetization of our Co particle in the low damping limit ($\alpha
= 0.1$) for a very small ellipsoidal shaped particle of length $L=4$nm
and diameter $L=2$nm.

The $z$ and the $x$ components of the magnetization are shown as well
as its absolute value.  The magnetic field ${\mathbf B}$ is set under
an angle of $45^{\circ}$ to the $z$ axis within the $yz$ plane so that
the response of the system to the external field sets in directly. The
wavering magnetization of the system clearly follows from the
precession of the spins.

Note, that the precession time of our system is not simply given by
the precession time of a single spin in an external field ($\tau_{p} =
2 \pi (1+ \alpha^2)/\gamma B \approx 32$ ps in our case). Instead, the
whole internal field is relevant for the spin precession, also the
contributions from the dipolar field, the exchange and the anisotropy.
This internal field is not constant in time and in general it is
non-homogenous within the system. However, in the very small particle
which we consider here, the internal field is sufficiently homogenous
so that the magnetization moves coherently. Hence, the absolute value
of the magnetization remains constant in time as is shown in the
figure.  Note also, that even after the new stable state is reached
the magnetization still keeps on oscillating around its equilibrium
value, driven by thermal fluctuations.

\section{Thermally Activated Reversal}

\begin{figure}
  \includegraphics[width=.85\textwidth, bb = 100 260 450
  470]{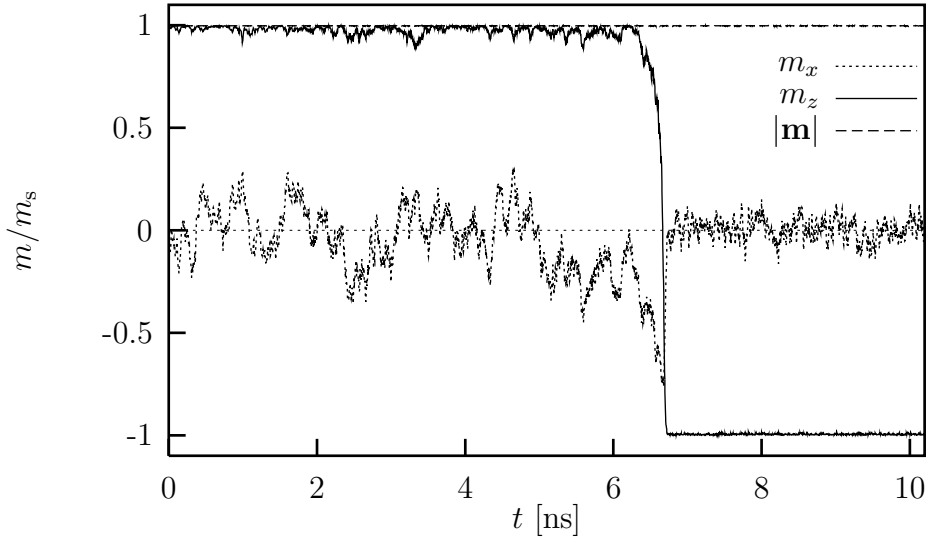} 
  \caption{Reduced magnetization vs. time for the same particle as in
    Fig. \ref{f:mag_LD}. The data are from Monte Carlo simulations. $B
    = 1.1$T and $T = 5$ K.}
\label{f:mag_MC}
\end{figure}

Let us now turn to the case $B < B_c$, where the reversal process can
only occur when it is thermally activated. Since we are now interested
in the long time and high damping limit ($\alpha = 4$) where the
behavior of the particle is governed by thermal fluctuations we can
use Monte Carlo methods.  Figure \ref{f:mag_MC} shows the typical time
dependence of the magnetization of the same Co particle as before. The
field $\mathbf{B}$ is set antiparallel to the initial state here, so
that the zero-temperature coercive field is given by $B_c =
2d_{\mathrm{eff}}V /\mu_{\mathrm s}$ which in our case is $B_c \approx
1.4$ T.

As one can see, the $z$ component of the magnetization remains nearly
constant for a time period which is rather long as compared to the
previous simulation.  Then, suddenly, the magnetization drops and its
sign changes.  As before the absolute value of the magnetization is
constant in time and the reversal mechanism is mainly a coherent
rotation.  The value of the switching time is approximately 6.6 ns in
our simulation.  However, this thermally activated switching is not a
deterministic process as it was the short time dynamics studied
before, where the switching followed mainly from the deterministic
part of the equation of motion.  Instead, the thermal activation
process here is a probabilistic event.  The probability distribution
$P(t_s)$ for switching events taking place after a time $t_s$ follows
an exponential law \cite{brownPR63},
\begin{equation}
P(t_s)  \sim \exp{(-t_{\mathrm s}/\tau)},
\label{e:ver}
\end{equation}
in the limit of large time-scales. Here, $\tau$ is a characteristic
time scale, 
\begin{equation}
\tau = \tau_{0} \exp{\big( \Delta E/ \kB T \big)},
\end{equation}
where $\tau_0$ is a prefactor and $\Delta E$ an energy barrier which
both are related to a certain reversal mechanism (see e.\ g.\ 
\cite{brownPR63,braunPRL93,coffeyPRL98} for analytically determined
prefactors and energy barriers in different systems and
\cite{nowakPRL00,hinzkePRB00,hinzkeJMMM00,nowakARCP00} for numerical
work on this subject). In general the prefactor may depend on the
system parameters, the temperature, the applied magnetic field and the
damping constant.

For the case of a Stoner-Wohlfarth particle with the applied field
parallel to the easy axis the energy barrier has the form
\begin{equation}
 \Delta E = d_{\mathrm{eff}} V \Big(1 - \frac{B}{B_c}\Big)^2.
 \label{e:energy}
\end{equation}
This energy barrier as well as the prefactor have been calculated by
Brown \cite{brownPR63} under the assumption that all magnetic moments
are parallel, so that the system behaves like one single magnetic
moment.

\begin{figure}
  \includegraphics[width=.85\textwidth, bb = 100 260 450
  470]{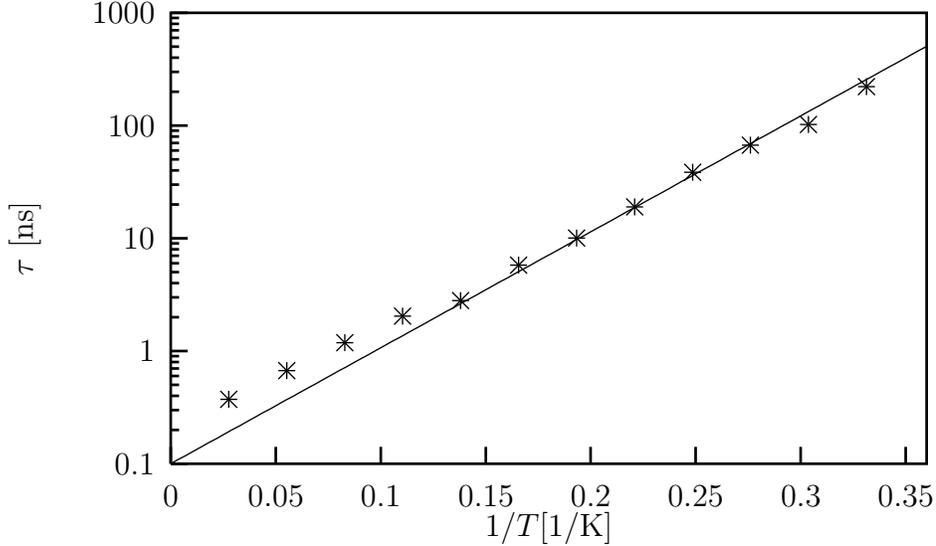} 
  \caption{Characteristic time $\tau$ vs. $1/ T$ for the same Co
    particle as in the figures before. The slope of the solid line
    represents the energy barrier $\Delta E$. $B = 1.1$ T.}
  \label{f:tau}
\end{figure}

For a further analysis we extract the energy barrier which governs the
reversal process from our numerical data.  Figure \ref{f:tau} shows
the temperature dependence of the characteristic time, i.\ e.\ the
mean switching time, obtained from our simulations.  The slope of the
solid line corresponds to the theoretical value of the energy barrier
for a reversal by coherent rotation obtained from Eq.~\ref{e:energy}.
Obviously, it is in very good agreement with our numerical data for
low enough temperatures.

\section*{Acknowledgments}
This work was supported by the Deutsche Forschungsgemeinschaft (SFB
491 and project NO290/1) and by the European Union (COST action P3,
working group 4).

\end{document}